\documentclass[twocolumn, iop, apj, twocolappendix, numberedappendix, appendixfloats]{emulateapj}
\usepackage{CJKutf8}
\usepackage{threeparttable}
\usepackage{multirow}
\usepackage{times}
\usepackage{enumitem}
\usepackage{url} 
\usepackage{color}
\usepackage{amsmath,bm}
\usepackage{subfigure}
\usepackage{natbib}
\usepackage{graphicx}
\usepackage{graphics}
\usepackage{hyperref}                             
\usepackage{amsmath}
\usepackage{bm}
\usepackage{subfigure}
\usepackage{array}
\usepackage{booktabs}
\usepackage{lineno}

\shorttitle{The Age-velocity Dispersion Relations}
\shortauthors{Sun et al.}
\submitted{Submitted to ApJ;\,\,\,\,Accepted December 09, 2024}

\begin{document}

%\linenumbers

\title{The Age-velocity Dispersion Relations of the Galactic Disk as Revealed by the LAMOST-Gaia Red Clump Stars}

\author{Weixiang Sun\textsuperscript{1,5}}
\author{Han Shen\textsuperscript{3,4}}
\author{Biwei Jiang\textsuperscript{1,5}}
\author{Xiaowei Liu\textsuperscript{2,5}}

\altaffiltext{1}{School of Physics and Astronomy, Beijing Normal University, Beijing 100875, People’s Republic of China; {\it sunweixiang@bnu.edu.cn {\rm (WXS)}}; {\it bjiang@bnu.edu.cn {\rm (BWJ)}}}
\altaffiltext{2}{South-Western Institute for Astronomy Research, Yunnan University, Kunming 650500, People's Republic of China; {\it x.liu@ynu.edu.cn {\rm (XWL)}}}
\altaffiltext{3}{School of Physics, University of New South Wales, Kensington 2032, Australia}
\altaffiltext{4}{ARC Centre of Excellence for All Sky Astrophysics in 3 Dimensions (ASTRO 3D), Australia}
\altaffiltext{5}{Corresponding authors}

\begin{abstract}

Using nearly 230,000 red clump (RC) stars selected from LAMOST and Gaia, we conduct a comprehensive analysis of the stellar age-velocity dispersion relations (AVRs) for various disk populations, within 5.0\,$\leq$\,$R$\,$\leq$\,15.0\,kpc and $|Z|$\,$\leq$\,3.0\,kpc.
The AVRs of the whole RC sample stars are accurately described as $\sigma_{v}$\,=\,$\sigma_{v,0}$\,($\tau$\,+\,0.1)$^{\beta_{v}}$, with $\beta_{R}$, $\beta_{\phi}$ and $\beta_{Z}$ displaying a global exponential decreasing trend with $R$, which may point to the difference in spatial distributions of various disk heating mechanisms.
The measurements of $\beta$ -- $R$ for various disks suggest that the thin disk exhibits a radial dependence, with a global exponential decreasing trend in $\beta_{R}$ -- $R$ and $\beta_{Z}$ --$R$, while $\beta_{\phi}$ remains a nearly constant value (around 0.20$\sim$0.25) within 8.5 $\leq$ $R$ $\leq$ 11.5\,kpc.
The thick disk displays a global increasing trend in $\beta_{R}$ -- $R$, $\beta_{\phi}$ -- $R$ and $\beta_{Z}$ -- $R$.
These results indicate that the thin disk stars are likely heated by long-term heating from GMCs and spiral arms, while thick disk stars are likely heated by some violent heating process from merger and accretion, and/or formed by the inside-out and upside-down star formation scenarios, and/or born in the chaotic mergers of gas-rich systems and/or turbulent ISM.
Our results also suggest that the disk perturbation by a recent minor merger from Sagittarius may have occurred within 3.0\,Gyr.

\end{abstract}


\section{Introduction}

The dynamic history of the Milky Way is imprinted in stellar kinematics.
Therefore, the measurement of stellar kinematics as a function of stellar ages over a larger disk volume can not only make a deep understanding of the nature of the disk, but also constrain the dynamic history of our galaxy \citep[e.g.,][]{Holmberg2009, Minchev2013}.
One of the hottest topics is the study of stellar age-velocity dispersion relations (AVRs) \citep[e.g.,][]{Stromberg1946, Casagrande2011, Sharma2014}.

The AVRs are widely confirmed to be a powerful tool for constraining the structures and evolution of the Milky Way \citep[e.g.,][]{Wielen1977, Nordstrom2004, Sharma2021}.
Those results indicate that the stellar velocity dispersions display a global increasing trend with age in the solar vicinity \citep[e.g.,][]{Aumer2009, Yu2018}.

Spitzer \& Schwarzschild ({\color{blue}{1951, 1953}}) first to demonstrate that Giant Molecular Clouds (GMCs) could contribute to the disk heating, while Barbanis \& Woltjer ({\color{blue}{1967}}) suggested similar effects from transient spiral arms.
These ideas have been further developed in several studies \citep[e.g,][]{Jenkins1990, Fuchs2001}.
Most of those results suggested that the slope of the vertical AVR in the solar neighborhood is well-modeled as $\sigma_{Z}$ $\propto$ $\tau^{\beta_{Z}}$ with $\beta_{Z}$ $\sim$ 0.25 \citep[e.g.,][]{Lacey1984, Hanninen2002}, while it is usually obviously smaller than observations \citep[where $\beta_{Z}$ is closer to 0.35--0.6, e.g.,][]{Wielen1977, Seabroke2007, Soubiran2008, Sun2024a}.
An increased contribution from GMCs heating during the early stage of disk evolution can be invoked to address these discrepancies \citep[e.g.,][]{Aumer2016a, Aumer2016b}, and a recent analysis based on APOGEE-Gaia data working on the vertical actions as a function of age in low-[$\alpha$/Fe] stars is likely to confirm this notion \citep[][]{Ting2019}.
In addition, some non-axisymmetric features, such as the well-known bars, warp, flare, as well as rich substructures, may also contribute to a part of the disk heating \citep[e.g.,][]{Grand2016, Mackereth2019, Sun2024a, Antoja2018}.

However, some observations tend to detect a jump in the AVRs for stars older than 7.0$\sim$9.0\,Gyr \citep[e.g.,][]{Quillen2001, Sun2024a}.
This is likely related to a thick disk component, and points to the disk heating may require some violent dynamical events \citep[][]{Jenkins1992}. 
High-resolution cosmological simulations revealed a possible violent heating agent, that is, the perturbations from mergers by relatively massive ($M$ $\geq$ 10$^{10}$\,M$_{\odot}$) satellite \citep[][]{Grand2016}.
This is particularly relevant since the Galaxy may be heated by the merger \citep[e.g.,][]{Villalobos2008, Minchev2014, Belokurov2018, Deason2018, Helmi2018, Kruijssen2019}.
Furthermore, accretion from satellite galaxies \citep{Abadi2003}, and the infall of misaligned gas \citep[e.g.,][]{Roskar2010, Sharma2012, Aumer2013}, may also play important roles in the disk violent heating.
However, the impact of these violent heating events on the dynamics of the Galactic disk remains to be fully explored.

Recent studies suggested it has been informative to look at how the AVRs change with stellar populations \citep[e.g.,][]{Yu2018, Mackereth2019, Sun2024a}.
Most of those indicate that the thin and thick disks display an obvious gap in AVRs in the solar neighborhood, which may be related to the significant difference in the heating, formation and evolution histories of the two disks \citep[][]{Mackereth2019, Sun2024a}.
However, a detailed analysis of the AVRs based on different stellar populations is still not yet well measured over a larger disk volume since the samples used in previous studies lack accurate measurement of the stellar age and kinematics \citep[e.g.,][]{Aumer2009, Yu2018}, which would enable a credible assessment of the disk heating, formation and evolution histories.
At present, the larger sample of red clump (RC) stars \citep[e.g.,][]{Huang2020, Wang2023} selected from LAMOST \citep[e.g.,][]{Deng2012, Cui2012, Liu2014, Yuan2015}, combined with stellar kinematic parameters from Gaia data \citep[e.g.,][]{Gaia Collaboration2023a, Gaia Collaboration2023b, Recio-Blanco2023}, presents an excellent opportunity to study this field.
Based on this sample, it is possible to conduct an exhaustive study of the AVRs in a larger disk volume, thereby constraining the disk heating and evolution histories.

This paper is structured as follows.
In Section\,2, we describe the data used in this paper,
and present our results and discussion in Section\,3. 
Finally, our main conclusions are summarized in Section\,4.

\begin{figure}[t]
\centering
\subfigure{
\includegraphics[width=8.3cm]{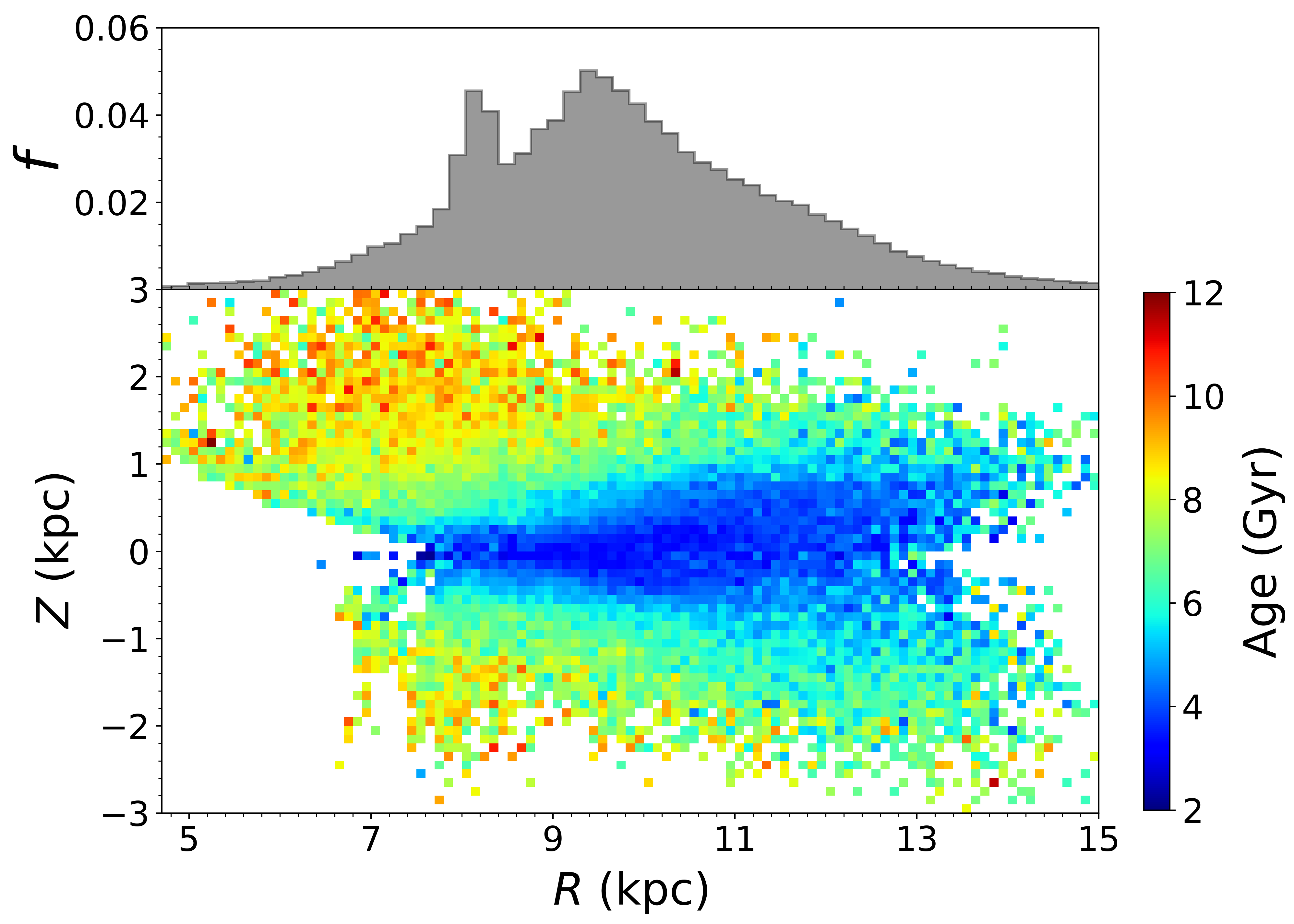}
}

\caption{Spatial distribution in the $R$ - $Z$ plane, of the sample stars, color-coded by the mean stellar ages.
There are no less than 8 stars in each bin, with spaced 0.1\,kpc in both axes.
The top panel displays histograms of fraction number density ($f = N_i/N_{\rm tot}$) distributions along the radial direction.
Here, $N_{i}$ represents the number of stars in each radial bin, and $N_{\rm tot}$ denotes the total number of stars in the whole sample.}
\end{figure}
%%\label{Fig.1}

\section{Data}

In this paper, we mainly used a sample with 228,820 RC stars selected Huang et al. ({\color{blue}{2020}}) and Wang et al. ({\color{blue}{2023}}) with selecting the sources with a high signal-to-noise ratio (S/N) for the common targets.
The typical uncertainties of the line-of-sight velocity ($V_{\rm r}$), effective temperature ($T_{\rm eff}$), surface gravity (log$_{g}$) [$\alpha$/Fe] and [Fe/H] are 5\,km s$^{-1}$, 100\,K, 0.10\,dex, 0.03$-$0.05\,dex and 0.10$-$0.15\,dex, respectively \citep[e.g.,][]{Xiang2017, Huang2020, Wang2023}.
Based on the standard-candle nature of these RC stars, their distance error is typically smaller than around 5\%--10\%.
To further improve the accuracy of the kinematic estimation, we update the atmospheric parameters of the whole RC stars to Gaia DR3 \citep[e.g.,][]{Gaia Collaboration2023a, Gaia Collaboration2023b, Recio-Blanco2023}.

The standard Galactocentric cylindrical Coordinate ($R$, $\phi$, $Z$) is used in this paper, with the three velocity components respectively $V_{R}$, $V_{\phi}$ and $V_{z}$.
To estimate the 3D positions and kinematics, we assume the Galactocentric distance of the Sun as $R_{\odot}$ = 8.34 kpc \citep{Reid2014}, the local circular velocity as $V_{c,0}$ = 238 km s$^{-1}$ \citep[e.g.,][]{Reid2004, Schonrich2010, Schonrich2012, Reid2014, Huang2016, Bland-Hawthorn2016} and the solar motions as ($U_{\odot}$, $V_{\odot}$, $W_{\odot}$) $=$ $(13.00, 12.24, 7.24)$ km s$^{-1}$ \citep{Schonrich2018}.

\begin{table*}
\caption{The properties of thin and thin disk populations.}
\centering
\setlength{\tabcolsep}{7mm}{
\resizebox{2.\columnwidth}{!}{
\begin{tabular}{lllllllll}
\hline
\hline
\specialrule{0em}{7pt}{0pt}
Name                                         &   $\left\langle R \right\rangle$   &   $\left\langle V_{\phi} \right\rangle$   &     $\sigma_{R}$        &    $\sigma_{\phi}$    &       $\sigma_{Z}$       &  $\left\langle V_{R}V_{Z} \right\rangle$  &     Number     &    n\,(ratio) \\[0.07cm]
                                             &                (kpc)               &               (km\,s$^{-1}$)              &     (km\,s$^{-1}$)      &     (km\,s$^{-1}$)    &      (km\,s$^{-1}$)      &          (km$^{2}$\,s$^{-2}$)             &                &                \\
\specialrule{0em}{7pt}{0pt}
\hline
\specialrule{0em}{7pt}{0pt}
Thin disk                                    &                 10.08              &                     224.82                &        34.59            &           22.88       &          19.72           &                19.51             &    128,878     &      80.67\%\\ [0.2cm]
Thick disk                                   &                 8.57               &                     176.94          &        65.47            &           54.07       &          41.32           &                115.91                     &    19,261      &      12.06\%   \\
\specialrule{0em}{7pt}{0pt}
\hline
\specialrule{0em}{7pt}{0pt}
\end{tabular}}
}
\label{tab:datasets}
\end{table*}

\begin{figure*}[t]
\centering

\subfigure{
\includegraphics[width=17.5cm]{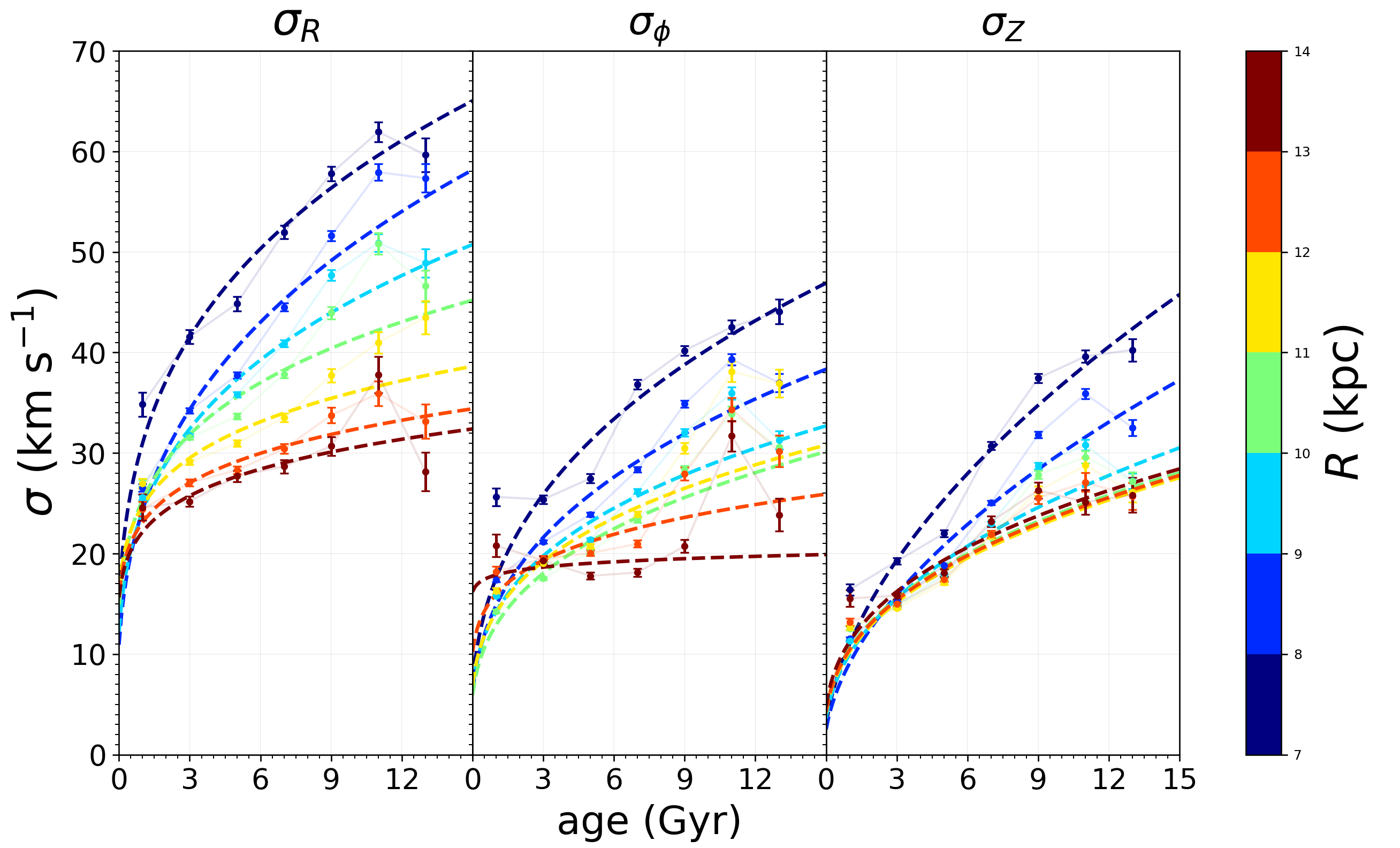}
}

\caption{The AVRs of the whole RC sample stars at different $R$ bins, color-coded by mean $R$.
From left to right represent respectively the AVR of $\sigma_{R}$, $\sigma_{\phi}$ and $\sigma_{Z}$.
The dashed lines represent the best fit with equation (1) for various $R$ bins.}
\end{figure*}
%%\label{Fig.2}

\begin{figure*}[t]
\centering

\subfigure{
\includegraphics[width=17.cm]{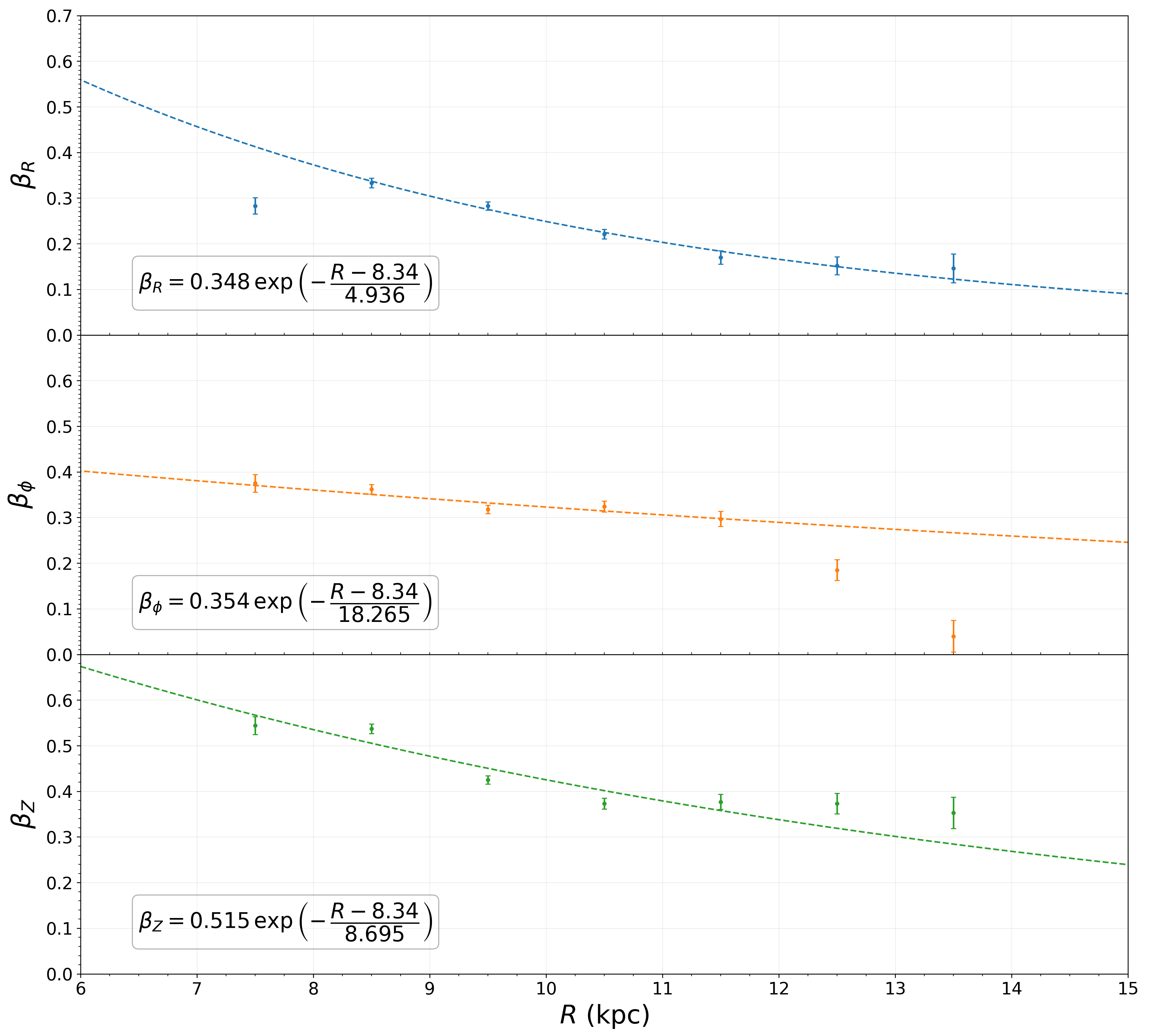}
}

\caption{The best fit parameter ($\beta$) of Equation (1) as a function of $R$ of the whole RC sample stars, with the $\beta_{R}$, $\beta_{\phi}$ and $\beta_{Z}$ are plotted by different colors.}
\end{figure*}
%%\label{Fig.3}

\begin{figure*}[t]
\centering

\subfigure{
\includegraphics[width=16.7cm]{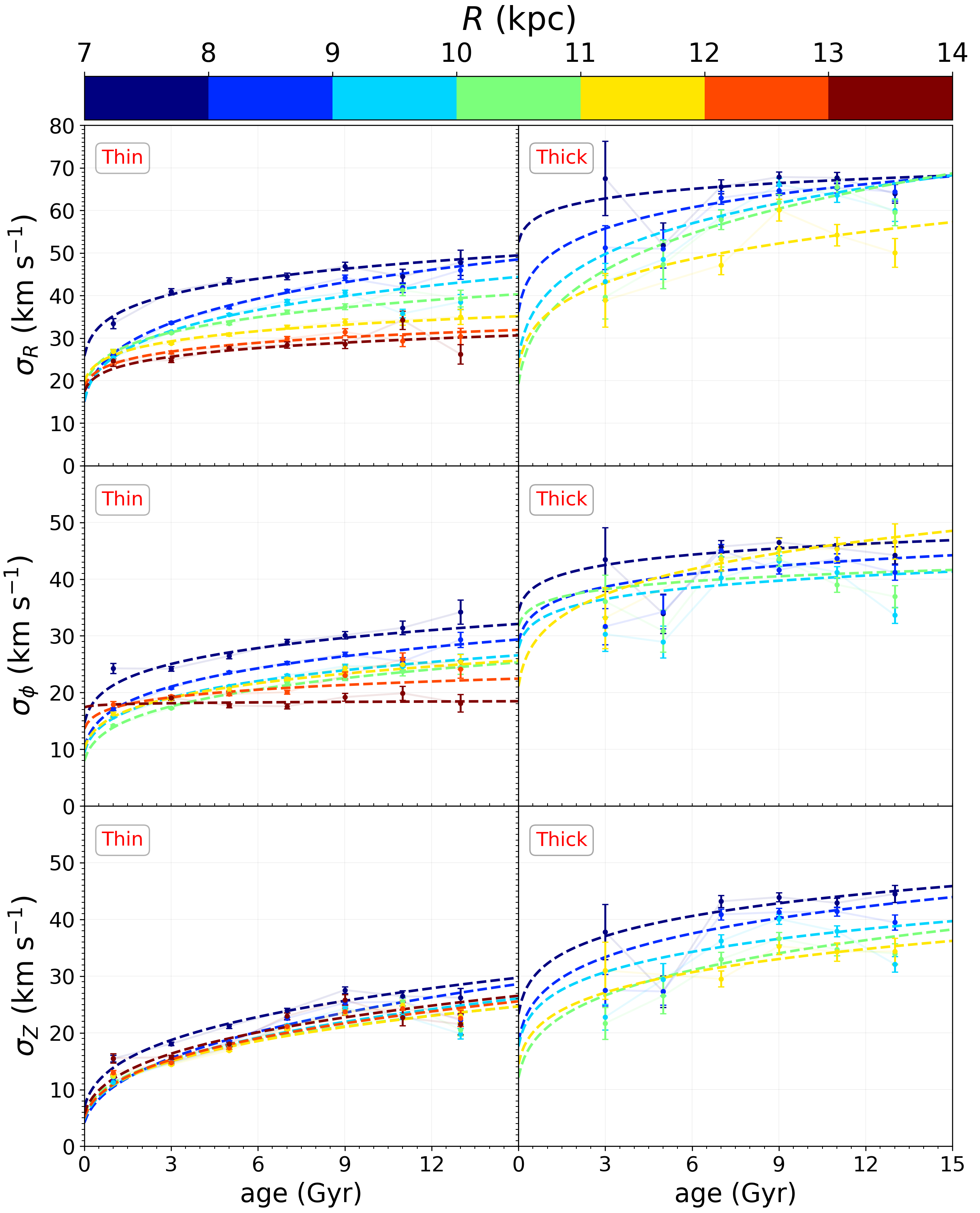}
}
\caption{Similar to Fig.\,2, but for the thin (left panel) and thick (right panel) disks.}
\end{figure*}
%%\label{Fig.4}

\begin{figure*}[t]
\centering

\subfigure{
\includegraphics[width=17.cm]{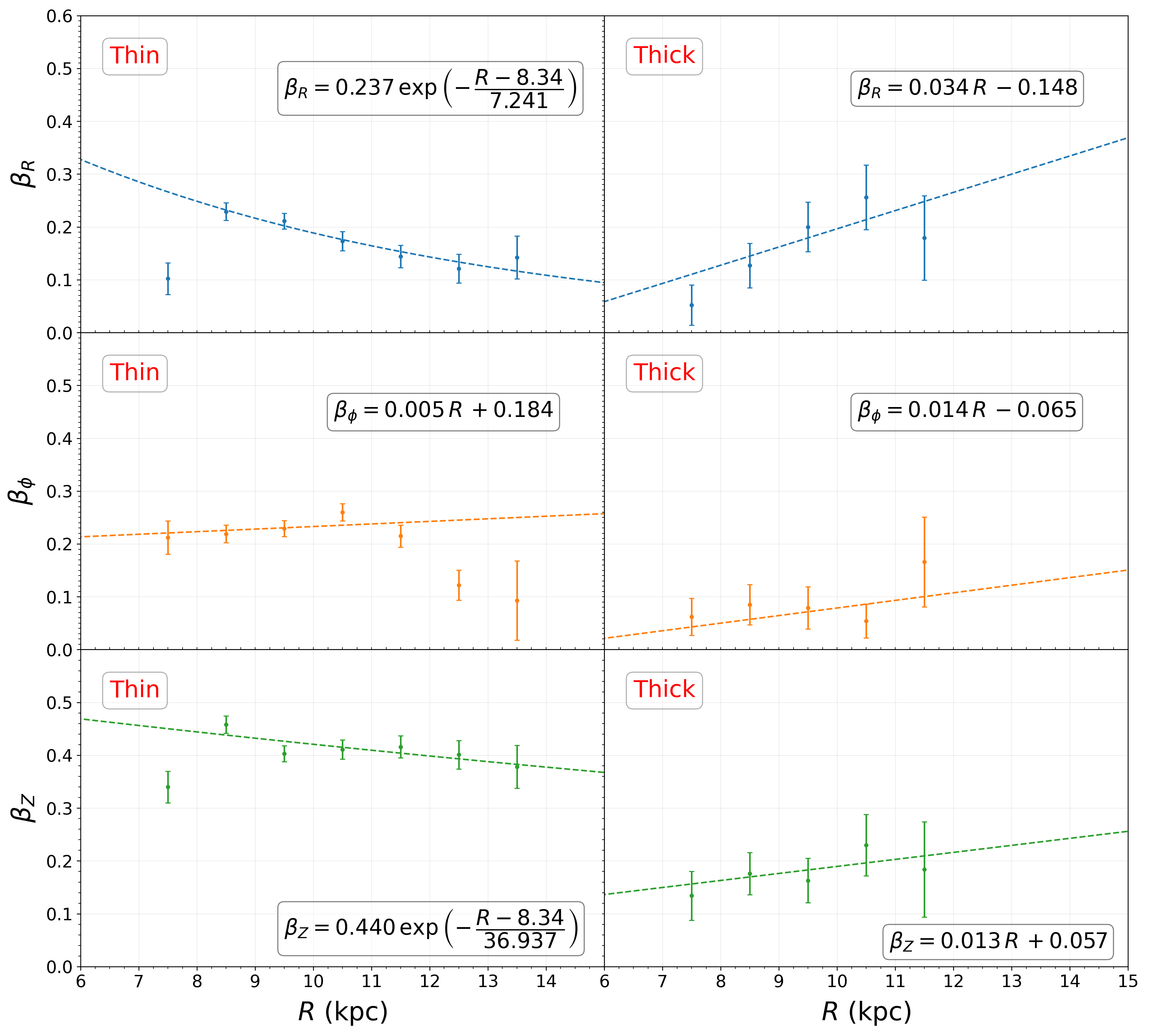}
}

\caption{Similar to Fig.\,3 but for the thin (left panel) and thick (right panel) disks.}
\end{figure*}
%%\label{Fig.5}

The velocity dispersions ($\sigma$) per bin are estimated by 3$\sigma$ clipping produced, and to further improve the accuracy of the stellar kinematics, we use cuts with stellar SNRs $>$ 20 and the distance error $\leq$ 10\%.
Those cuts can ensure the determined uncertainties of the 3D velocities are typical within 5.0\,km s$^{-1}$.
We also adopt the cuts with $\ |V_{z}|$ $\leq$ 120 km s$^{-1}$ and [Fe/H] $\geq -1.0$ dex to exclude any possible halo stars \citep{Huang2018, Hayden2020}.
We also removed the so-called ``young" [$\alpha$/Fe]-enhanced stars with age $\leq$ 6.0\,Gyr and [$\alpha$/Fe] $\geq$ 0.15\,dex since their true ages are confirmed have been underestimated, that is caused by the present large masses result from the binary merger or mass transfer \citep[e.g.,][]{Sun2020}.
With the above conditions, we finally selected 159,752 RC stars, and the stellar age distribution of the final RC sample stars in the $R$--$Z$ plane is shown in Fig.\,1.

To gain a deep understanding of the disk heating and evolution histories from the AVRs, we further separate our sample into thin and thick disks on the [Fe/H]--[$\alpha$/Fe] plane similar to Sun et al. ({\color{blue}{2023, 2024a}}), and finally, 19,261 stars and 128,878 stars are selected for the thick and thin disks, respectively.
The properties of the two disks are shown in Table\,1.

\section{Results and Discussion}

\subsection{The AVRs of the whole RC sample stars}

The AVRs of the whole RC sample stars are shown in Fig.\,2, color-coded by stellar $R$.
The panels from left to right represent the AVR of velocity in radial, azimuthal and vertical directions, respectively.

The velocity dispersion ($\sigma$) of the whole RC sample displays an obvious global trend with stellar age and $R$, which increase with stellar age and decrease with $R$ (see Fig.\,2).
In the solar neighbourhood (8.0 $\leq$ $R$ $<$ 9.0\,kpc), the $\sigma_{R}$, $\sigma_{\phi}$ and $\sigma_{Z}$ increases steadily from, respectively, 34.0\,km\,s$^{-1}$, 20.0\,km\,s$^{-1}$ and 15.0\,km\,s$^{-1}$ at age $\sim$ 3.0\,Gyr to, respectively, 50.0\,km\,s$^{-1}$, 35.0\,km\,s$^{-1}$ and 32.0\,km\,s$^{-1}$ at age larger than around 9.0\,Gyr.
This result is in good agreement with previous studies in the solar neighborhood \citep[e.g.,][]{Yu2018, Sun2024a}.
In the outer disk regions ($R$ $\geq$ 10.0\,kpc), the AVRs in $\sigma_{Z}$ show one $R$ bin intersect another, which may be related to the well-known disk flare nature in the outer Galactic disk \citep[e.g.,][]{Minchev2015, Mackereth2017, Sanders2018}.
In addition, the significant warp of the out disk \citep[e.g.,][]{Jonsson2024, Uppal2024, Khanna2024, Liu2024, Sun2024d} and the disturbance in the outer disk \citep[e.g.,][]{McMillan2022}, may also contribute to this behavior.

The AVR profile of the whole RC sample displays significant variations across various $R$ bins, with the trend flattening out as $R$ increases (see Fig.\,2).
We fitted these profiles with a simple power law as follows:
\begin{equation}
\sigma_{v} = \sigma_{v,0} (\tau + 0.1)^{\beta_{v}}
\end{equation}
Here we set velocity dispersion ($\sigma_{v}$) changes with ($\tau + 0.1)^{\beta_{v}}$ since the stellar birth velocity dispersion contributes AVRs is typically considered to be no less than 0.1\,Gyr \citep[e.g.,][]{Sharma2021}.
The $\sigma_{v,0}$ is the constant of the exponential growth of $\sigma_{v}$ with age ($\tau$).

The best fits of AVRs are plotted by dashed lines in Fig.\,2, and the best-fit parameters ($\beta$) as a function of $R$ are shown in Fig.\,3.
The $\beta_{R}$, $\beta_{\phi}$ and $\beta_{Z}$ display a global decreasing trend as $R$ increases,
with the $\beta_{R}$, $\beta_{\phi}$ and $\beta_{Z}$ decreases steadily from, respectively, 0.33, 0.36 and 0.54 at $R$ $\sim$ 8.5\,kpc to, respectively, 0.17, 0.30 and 0.39 at $R$ larger than around 11.5\,kpc.
We can also observe that the trend of $\beta_{Z}$ -- $R$ becomes noticeably flat at $R$ $\geq$ 10.5\,kpc, which is likely also related to disk flare in the outer Galactic disk.
The results also indicate that a global trend with $\beta_{R}$ $<$ $\beta_{\phi}$ $<$ $\beta_{Z}$ at the same $R$.
This global trend is also confirmed in the solar neighborhood by previous studies \citep[e.g.,][]{Yu2018, Sun2024a}.

In the solar neighborhood (8.0 $\leq$ $R$ $<$ 9.0\,kpc), our results indicate that $\beta_{R}$ = 0.333 $\pm$ 0.011, $\beta_{\phi}$= 0.362 $\pm$ 0.010, and $\beta_{Z}$= 0.537 $\pm$ 0.013, which are in good agreement with previous observations with $\beta_{R}$  = 0.19 $\sim$ 0.35 and $\beta_{Z}$  = 0.35 $\sim$ 0.6  \citep[e.g.,][]{Aumer2009, Sharma2014, Sun2024a}, while being slightly larger than that of the prediction from simulations of GMCs with $\beta_{R}$ = 0.21 $\pm$ 0.02, $\beta_{Z}$ = 0.26 $\pm$ 0.03 \citep[e.g.,][]{Lacey1984, Hanninen2002}.
The reason for this discrepancy may be clarified by our results in Fig.\,2.
The results display all AVRs have a significant jump increasing at age around 7.0$\sim$9.0\,Gyr, which may be related to a thick disk component that is caused by some violent heating from merger \citep[e.g.,][]{Quinn1993, Grand2016}, accretion \citep{Abadi2003}, and the infall of misaligned gas \citep[e.g.,][]{Roskar2010, Sharma2012}, and/or is born in the chaotic mergers of gas-rich systems and/or turbulent interstellar medium \citep[ISM; e.g.,][]{Brook2004, Brook2007, Brook2012, Wisnioski2015}.

The global trends of $\beta$ -- $R$ in our results are obviously (see Fig.\,3), with all profiles break at $R$ $\sim$ 8.5\,kpc, with additional $\beta_{\phi}$ -- $R$ profile also break at $R$ $\sim$ 11.5\,kpc.
Since the number of stars in the bin of $R$ = [7.0, 8.0]\,kpc is larger than 8,000 stars, it can ensure the statistical reliability, and therefore, the $R_{break}$ $\sim$ 8.5\,kpc may be caused by the stars in inner disk region trending to be distributed at large vertical Galactic heights (see Fig.\,1).
The $R_{break}$ $\sim$ 11.5\,kpc of $\beta_{\phi}$ -- $R$ we consider that may be related to a recent disk heating event.
Support for this hypothesize comes from the AVRs in $\sigma_{\phi}$ at $R$ = [12.0, 13.0]\,kpc and $R$ = [13.0, 14.0]\,kpc, which display the stars with age range from $\sim$3.0\,Gyr to $\sim$7.0\,Gyr have a similar value in $\sigma_{\phi}$, meaning those stars are likely to be rapidly heated to such large azimuthal velocity dispersion in a short time-scale by a violent event.
Considering that the $\sigma_{\phi}$ of stars with age younger than $\sim$3.0\,Gyr obviously larger than that of stars with age = 6.0$\sim$7.0\,Gyr in these $R$ bins, we suggest this violent event may have occurred within 3.0\,Gyr, which may be the minor merger/perturbation from the Sagittarius passing by \citep[e.g.,][]{Gomez2012a, Gomez2012b, Laporte2018, Carrillo2019}

At 8.5 $\leq$ $R$ $\leq$ 11.5\,kpc, the $\beta_{R}$ -- $R$ and $\beta_{Z}$ -- $R$ have a similar gradient, which is slightly stronger than the gradient of $\beta_{\phi}$ -- $R$, meaning that the $\beta_{R}$ and $\beta_{Z}$ are sensitive with $R$ than that of $\beta_{\phi}$, which may be related to the difference in spatial distributions of various disk heating mechanisms.

We fit the profiles of $\beta$ -- $R$ with a exponential function:
\begin{equation}
\beta_{v}(R) = \beta_{v,0} \,\operatorname{exp}\left( - \frac{R - R_{0}}{R_{v}} \right)
\end{equation}
Here, we defined the $\beta_{v}$ does exponential decrease with $R$, where $\beta_{v,0}$ represent the value of the $\beta_v$ at solar radius ($R$ = 8.34\,kpc), and the $R_{v}$ represent the scale length of the exponential decrease of $\beta_v$.

Considering the $R_{break}$ in AVRs and the Sagittarius perturbation on the contribution of $\beta_{\phi}$ -- $R$ at $R$ $\geq$ 11.5\,kpc we discussed above, we only fit the profiles of $\beta_{R}$ -- $R$ and $\beta_{Z}$ --$R$ at $R$ $\geq$ 8.5\,kpc, and fit the profile of $\beta_{\phi}$ -- $R$ at 8.5 $\leq$ $R$ $\leq$ 11.5\,kpc.
The results of the best fits are plotted with dashed lines as shown in Fig.\,3.

The result yields $\beta_{R}$ = 0.348 exp ($-$($R$\,$-$\,8.34)$/$4.936), $\beta_{\phi}$ = 0.354 exp ($-$($R$\,$-$\,8.34)$/$18.265) and $\beta_{Z}$ = 0.515 exp ($-$($R$\,$-$\,8.34)$/$8.695).
Based on the best fits, we can accurately determine the $\beta_{R}$, $\beta_{\phi}$and $\beta_{Z}$ at the solar radius ($R$ = 8.34\,kpc) are $\beta_{R,0}$ = 0.348 $\pm$ 0.007, $\beta_{\phi,0}$ = 0.354 $\pm$ 0.005 and $\beta_{Z,0}$ = 0.515 $\pm$ 0.007, which are in good agreement with the previous observations in the solar neighborhood \citep[e.g.,][]{Aumer2009, Sharma2014, Sun2024a}, while is also slightly larger than that of the prediction from simulations of the scattering by GMCs \citep[e.g.,][]{Lacey1984, Hanninen2002}.

\subsection{The AVRs of the thin and thick disks}

The fact that the increasing trend in AVRs is the contribution of Galactic dynamical heating, from the scattering of GMCs and spiral arms, and the resonance with the Galactic bar \citep[e.g.,][]{Kalnajs1991, Dehnen1998, Dehnen2000, Antoja2014, Sun2024c}, as well as the perturbation of disk sub-structures \citep[e.g.,][]{Antoja2018, Hunt2018}.
Most of these heating processes result in a simple power-law increase in velocity dispersions with age, that is $\sigma_{v}$ $\propto$ $\tau^{\beta_v}$ \citep[e.g.,][]{ Lacey1984, Hanninen2002}.
Therefore, the AVRs display a jump increase at age = 7.0$\sim$9.0\,Gyr (see Fig.\,2) is a surprising result.
Similar result is also detected by previous observations in the solar neighborhood, which suggested that this jump is likely related to a thick disk component caused by merger events \citep[][]{Quillen2001, House2011}, while other studies tend to suggest a simple power-law profile in AVRs that shaped by long-term heating \citep[][]{Aumer2009, Sharma2021}.

To further explore those debates, we present the AVRs in terms of chemically thin and thick disks in Fig.\,4, and plot $\beta$ -- $R$ in Fig.\,5.
The thin and thick disks at the same $R$ bin display a significant gap in AVRs (see Fig.\,4), which indicates that the youngest thick disk stars with age = 3.0$\sim$6.0\,Gyr typically exhibit larger velocity dispersions than the oldest thin disk stars with age = 9.0$\sim$13.0\,Gyr, suggesting an obvious difference in the heating histories of the thin and thick disks. 
Similar behavior has also been reported by previous studies in the solar neighborhood \cite[e.g.,][]{Sun2024a}.
It is worth noting that the jump in AVRs at age = 7.0$\sim$9.0\,Gyr has almost disappeared in the thin and thick disks results (see Fig.\,4), indicating that this jump in AVRs of the whole sample stars most likely points to the presence of a thick disk component that shaped by some violent heating processes.
In addition, in the outer disk regions ($R$ $\geq$ 10.0\,kpc), the AVRs of $\sigma_{Z}$ for thin disk stars show one $R$ bin intersects another, while the thick disk displays no such obvious behavior, which indicates that the flare in the outer disk is a unique property of the thin disk \citep[e.g.,][]{Vickers2021, Sun2024b, Khanna2024}.

In Fig.\,5, the $\beta_{R}$ and $\beta_{Z}$ of the thin disk also display a global decreasing trend with $R$, in the Galactic anti-center direction.
The $\beta_{R}$ decreases steadily from 0.23 at $R$ $\sim$ 8.5\,kpc to 0.12 at R larger than around 12.5\,kpc, and the $\beta_{Z}$ slightly decreases from 0.46 at $R$ $\sim$ 8.5\,kpc to 0.40 at R larger than around 12.5\,kpc.
Apart from the fact that a rapid decrease in $\beta_{\phi}$ caused by a recently minor merger by Sagittarius at $R$ $>$ 11.5\,kpc can still be found, the $\beta_{\phi}$ of the thin disk seem to display no obvious dependence on $R$, with $\beta_{\phi}$ shows nearly constant value with around 0.20$\sim$0.25.
These values are in rough agreement with the prediction by long-term heating from GMCs \citep[][]{Spitzer1953, Jenkins1992} and spiral arms \citep[][]{Bissantz2003, Combes2014}.

The $\beta_{R}$ of the thick disk shows an obviously increasing trend with $R$, it increases steadily from 0.05 at $R$ $\sim$ 7.5\,kpc to 0.25 at R $\sim$ 11.5\,kpc.
However, the $\beta_{\phi}$ and $\beta_{Z}$ of the thick disk display a global weak increasing trend (or an almost constant value, that is $\beta_{\phi}$ = 0.05$\sim$0.15 and $\beta_{Z}$ = 0.15$\sim$0.20) with $R$.
These results combined with the global large velocity dispersions (see Fig.\,4) of the thick disk stars indicate that these stars are likely heated by some violent heating process by merger and accretion \citep[e.g.,][]{Kazantzidis2008, Belokurov2018, Kruijssen2019}.
The increasing trends in $\beta$ -- $R$ of the thick disk stars may indicate that the special star formation mechanism of these stars, such as, the ``inside-out" and ``upside-down" star formation scenarios \citep[e.g.,][]{Kawata2017, Schonrich2017, Bird2021}.
In addition, the globally small value of $\beta$ for the thick disk stars indicates that these stars, at all ages, display almost similar large velocity dispersions (a similar result is observed in the AVRs shown in Fig.\,4).
Although we cannot fully rule out the effectiveness of the AVRs, this finding also suggests that these stars are likely to have been rapidly heated to such large velocity dispersions in a short time-scale \citep[][]{Quinn1993, Abadi2003}, or are likely born in the chaotic mergers of gas-rich systems and/or turbulent ISM \citep[e.g.,][]{Brook2004, Brook2007, Wisnioski2015}.

The global profiles can also be clearly found in $\beta_{R}$ -- $R$ for the thin and thick disks within 8.5 $\leq$ $R$ $\leq$ 11.5\,kpc (see Fig.\,5), while only the $\beta_{R}$ -- $R$ and $\beta_{Z}$ -- $R$ profiles of the thin disk exhibit obvious exponential decreases, following Equation (2), others tend to display linear function shapes, and therefore, we fit the $\beta_{R}$ -- $R$ and $\beta_{Z}$ -- $R$ profiles of the thin disk using Equation (2), and fit other profiles with a linear function.
The best-fit results are shown with dashed lines in Fig.\,5.

The results yield the thin disk are $\beta_{R}$ = 0.237\,exp\,($-$($R$\,$-$\,8.34)$/$7.241), $\beta_{\phi}$ = 0.005\,$R$\,$+$\,0.184, and $\beta_{Z}$ = 0.440\,exp\,($-$($R$\,$-$\,8.34)$/$36.937).
The thick disk are $\beta_{R}$ = 0.034\,$R$\,$-$\,0.148, $\beta_{\phi}$ = 0.014\,$R$\,$-$\,0.065, and $\beta_{Z}$ = 0.013\,$R$\,$+$\,0.057.
Based on the best-fit results, we can accurately determine the $\beta_{R}$, $\beta_{\phi}$ and $\beta_{Z}$ of the thin disk stars at the solar radius ($R$ = 8.34\,kpc) are $\beta_{R,0}$ = 0.237 $\pm$ 0.013, $\beta_{\phi,0}$ = 0.226 $\pm$ 0.011 and $\beta_{Z,0}$ = 0.440 $\pm$ 0.015.
The thick disk stars at the solar radius are $\beta_{R,0}$ = 0.136 $\pm$ 0.015, $\beta_{\phi,0}$ = 0.052 $\pm$ 0.013 and $\beta_{Z,0}$ = 0.165 $\pm$ 0.016.
Our thin disk results are in good agreement with the previous observations in the solar neighborhood \citep[e.g.,][]{Yu2018, Sun2024a}, and align well with predictions from simulations of the scattering by GMCs \citep[e.g.,][]{Lacey1984, Hanninen2002} and spiral arms \citep[e.g.,][]{De Simone2004}.

\section{Conclusions}

In this paper, we used a sample with nearly 230,000 RC stars selected from LAMOST surveys to make a detailed analysis of the AVRs for different disk populations, in a large disk volume (5.0 $\leq$ $R$ $\leq$ 15.0\,kpc, and $|Z|$ $\leq$ 3.0\,kpc).
We find that:
\\
\\
$\bullet$ The AVRs of the whole RC sample stars can be well described as $\sigma_{v}$ = $\sigma_{v,0}$\,($\tau$\,+\,0.1)$^{\beta_{v}}$ with the $\beta_{v}$ decreases as $R$ increases, and exhibiting $\beta_{R}$ = 0.348 exp\,($-$($R$\,$-$\,8.34)$/$4.936), $\beta_{\phi}$ = 0.354 exp\,($-$($R$\,$-$\,8.34)$/$18.265) and $\beta_{Z}$ = 0.515\,exp\,($-$($R$\,$-$\,8.34)$/$8.695).
All AVRs of the whole RC sample stars have an obvious jump increasing at around age = 7.0$\sim$9.0 Gyr.
\\
\\
$\bullet$ The thin disk exhibits a radial dependence, with a global exponential decreasing trend in $\beta_{R}$ -- $R$ and $\beta_{Z}$ -- $R$.
The $\beta_{R}$ decreases steadily from 0.23 at $R$ $\sim$ 8.5\,kpc to 0.12 at R larger than around 12.5\,kpc, and following $\beta_{R}$ = 0.237 exp\,($-$($R$\,$-$\,8.34)$/$7.241).
The $\beta_{Z}$ decreases slightly from 0.46 at $R$ $\sim$ 8.5\,kpc to 0.40 at R larger than around 12.5\,kpc, and following $\beta_{Z}$ = 0.440 exp\,($-$($R$\,$-$\,8.34)$/$36.937).
The $\beta_{\phi}$ of the thin disk seem to display no obvious dependence on $R$ at 8.5 $\leq$ $R$ $\leq$ 11.5kpc, with $\beta_{\phi}$ = 0.20$\sim$0.25.
\\
\\
$\bullet$ The thick disk shows an obvious increasing trend in $\beta_{R}$ -- $R$, increasing steadily from $\beta_{R}$ = 0.05 at $R$ $\sim$ 7.5\,kpc to $\beta_{R}$ = 0.25 at R $\sim$ 11.5\,kpc, and following $\beta_{R}$ = 0.034\,$R$\,$-$\,0.148.
The $\beta_{\phi}$ and $\beta_{Z}$ display a global weak increasing trend (or an almost constant value, that is, $\beta_{\phi}$ = 0.05$\sim$0.15 and $\beta_{Z}$ = 0.15$\sim$0.20) with $R$, and exhibiting $\beta_{\phi}$ = 0.014\,$R$\,$-$\,0.065 and $\beta_{Z}$ = 0.013\,$R$\,$+$\,0.057.
\\
\\
These results indicate that the thin disk stars are likely heated by long-term heating from GMCs and spiral arms, while thick disk stars are likely heated by some violent heating process from merger and accretion, and/or formed by the inside-out and upside-down star formation scenarios, and/or born in the chaotic mergers of gas-rich systems and/or turbulent ISM.
In addition, the signal of the disk perturbation by a recent minor merger from Sagittarius can be also detected in our result, and the result suggests this disk perturbation event may have occurred within 3.0\,Gyr.

\section*{Acknowledgements}
We thank the anonymous referee for very useful suggestions to improve the work.
This work is supported by the NSFC projects 12133002, 11833006, and 11811530289, and the National Key R\&D Program of China No. 2019YFA0405500, 2019YFA0405503, and CMS-CSST-2021-A09, and the Postdoctoral Fellowship Program of CPSF under Grant Number GZC20240125, and the China Postdoctoral Science Foundation under Grant Number 2024M760240.

Guoshoujing Telescope (the Large Sky Area Multi-Object Fiber Spectroscopic Telescope LAMOST) is a National Major Scientific Project built by the Chinese Academy of Sciences. Funding for the project has been provided by the National Development and Reform Commission. LAMOST is operated and managed by the National Astronomical Observatories, Chinese Academy of Sciences.

\bibliographystyle{aasjournal}

\end{document}